\documentclass[prl, twocolumn, superscriptaddress]{revtex4}

\usepackage{graphicx} \usepackage{amssymb} \usepackage[usenames]{color}

\usepackage{amsmath}
\usepackage{xspace}

\begin{document}

\title{Isotope effect in quasi-two-dimensional metal-organic antiferromagnets} 
\author{P. A. Goddard}
\email{p.goddard1@physics.ox.ax.uk}
\affiliation{University of Oxford Department of Physics, 
Clarendon Laboratory, Parks Road,
Oxford OX1~3PU, U.K.}
\author{J. Singleton}
\affiliation{National High Magnetic Field Laboratory, 
Los Alamos National Laboratory, MS-E536, Los Alamos, NM 87545, USA}
\author{C. Maitland}
\affiliation{University of Oxford Department of Physics, 
Clarendon Laboratory, Parks Road,
Oxford OX1~3PU, U.K.}
\author{S. J.~Blundell}
\affiliation{University of Oxford Department of Physics, 
Clarendon Laboratory, Parks Road,
Oxford OX1~3PU, U.K.}
\author{T.~Lancaster}
\affiliation{University of Oxford Department of Physics, 
Clarendon Laboratory, Parks Road,
Oxford OX1~3PU, U.K.}
\author{P.~J.~Baker}
\affiliation{University of Oxford Department of Physics, 
Clarendon Laboratory, Parks Road,
Oxford OX1~3PU, U.K.}
\author{R.~D.~McDonald}
\affiliation{National High Magnetic Field Laboratory, 
Los Alamos National Laboratory, MS-E536, Los Alamos, NM 87545, USA}
\author{S.~Cox} 
\affiliation{National High Magnetic Field Laboratory, 
Los Alamos National Laboratory, MS-E536, Los Alamos, NM 87545, USA}
\author{P. Sengupta}
\affiliation{National High Magnetic Field Laboratory, 
Los Alamos National Laboratory, MS-E536, Los Alamos, NM 87545, USA}
\affiliation{Theoretical Division, Los Alamos National Laboratory, 
Los Alamos, NM 87545, USA}
\author{J. L. Manson}
\affiliation{Department of Chemistry and Biochemistry, 
Eastern Washington University, Cheney, WA 99004, USA}
\author{K. A. Funk}
\affiliation{Materials Science Division, 
Argonne National Laboratory, Argonne, IL 60439, USA}
\author{J. A. Schlueter}
\affiliation{Materials Science Division, 
Argonne National Laboratory, Argonne, IL 60439, USA}

\begin{abstract}
Although the isotope effect in superconducting materials is well-documented, changes in the magnetic properties of antiferromagnets due to isotopic substitution are seldom discussed and remain poorly understood. This is perhaps surprising given the possible link between the quasi-two-dimensional (Q2D) antiferromagnetic and superconducting phases of the layered cuprates. Here we report the experimental observation of shifts in the N\'{e}el temperature and critical magnetic fields ($\Delta T_{\rm N}/T_{\rm N}\approx 4\%$; $\Delta B_{\rm c}/B_{\rm c}\approx 4\%$) in a Q2D organic molecular antiferromagnets on substitution of hydrogen for deuterium. These compounds are characterized by strong hydrogen bonds through which the dominant superexchange is mediated. We evaluate how the in-plane and inter-plane exchange energies evolve as the hydrogens on different ligands are substituted, and suggest a possible mechanism for this effect in terms of the relative exchange efficiency of hydrogen and deuterium bonds.
\end{abstract}
\maketitle

In 1950 the isotope effect was first observed in the superconducting critical temperature of mercury~\cite{maxwell,reynolds}.
The effect was key to  verifying the electron-phonon pairing mechanism responsible for superconductivity in conventional superconductors and played an important role in motivating the microscopic theory of Bardeen, Cooper and Schrieffer~\cite{bcs}. The analogous pairing mechanism in the superconducting cuprates is still open to vigorous debate. It seems increasingly likely that some kind of magnetic interaction is responsible, arising from the underlying quasi-two-dimensional (Q2D) Heisenberg antiferromagnetic lattice~\cite{manousakis,monthoux91,monthoux94,moriya,schrieffer,harrison}. Nevertheless, an electron-phonon coupling component cannot yet be ruled out~\cite{nunner,sandvik,honerkamp}. One reason for this is that an adequate explanation of the anomalous, doping-dependent isotope effect observed in the cuprates is not forthcoming~\cite{franck,greco}. Given these considerations the importance of achieving an understanding of the effects of isotopic substitution on the properties of magnetic systems of reduced dimensionality is clear.

Metal-organic coordination complexes (see for example~\cite{jamie1,jamie1b,jamiechemcomm}) are an ideal framework in which to make studies of Q2D magnetism, and hence test theories relating to unconventional superconductivity. The extremely anisotropic Heisenberg exchange interactions observed in these materials are highly susceptible to small perturbations~\cite{goddardOMM}. Furthermore, the versatility of coordination chemistry allows for a high level of control over their composition and crystal structure. In consequence, the dimensionality of their magnetic structures can be readily changed via adjustments of molecular architecture and the magnitude of the exchange energies altered so that they are accessible by laboratory magnetic fields at cryogenic temperatures~\cite{goddardOMM}. Isotopic substitution provides an additional fine-scale tuning of magnetic properties. 

We have chosen to measure the coordination complex CuF$_2$(pyz)(H$_2$O)$_2$ (pyz =  C$_4$H$_4$N$_2$) because its quasi-two-dimensional antiferromagnetic properties are directly analogous to the parent compounds of the cuprate superconductors, which are similarly based on square lattices of Cu$^{2+}$ ions. We observe for the first time in such a system measurable isotope shifts in both the N\'{e}el temperature and the critical magnetic field.

\begin{figure}[t]
\centering
\includegraphics[width=4.7cm]{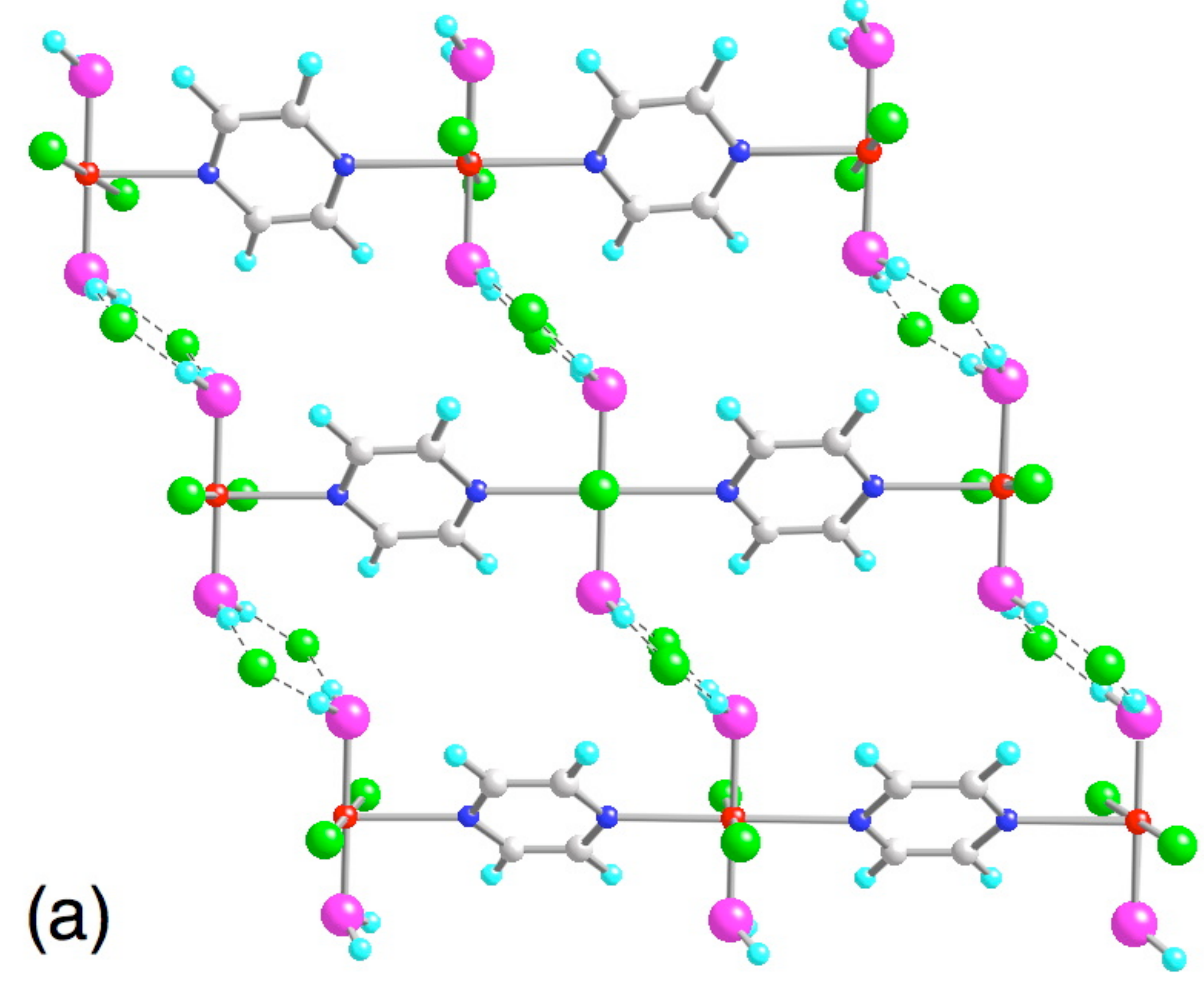}
\includegraphics[width=3.8cm]{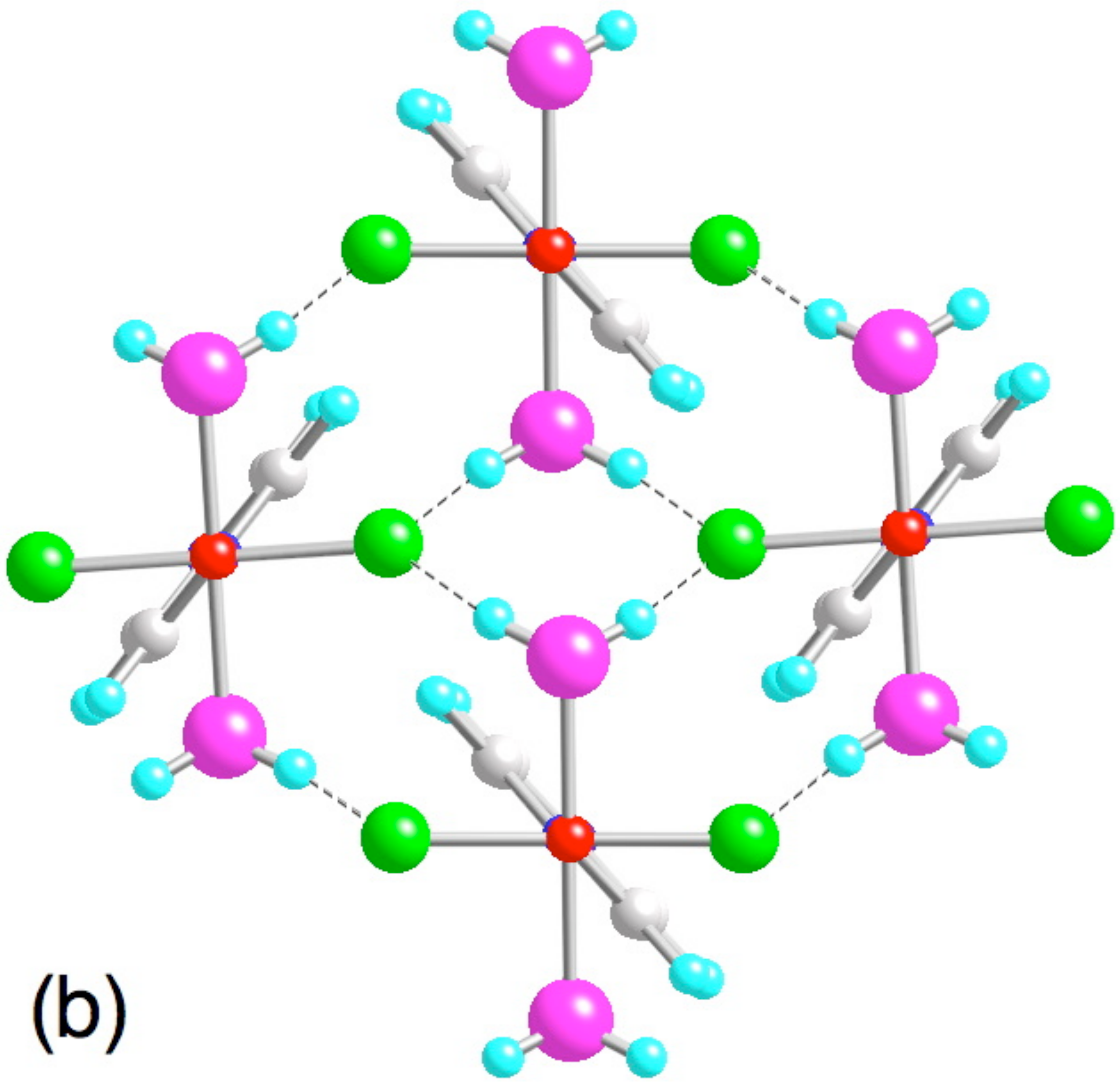}
\caption{Structure of CuF$_2$(pyz)(H$_2$O)$_2$ at room temperature viewed (a) perpendicular to the $a$-axis, showing the Cu-pyz-Cu chains; and (b) parallel to $a$-axis, showing the 2D magnetic planes and the F-H-O hydrogen bonds that mediate the dominant exchange interactions. Cu = red; F = green; N = blue; C = grey; H = cyan; and O = magenta. Hydrogen bonds are shown as dotted lines.}
\label{fig1}
\end{figure}

The structure of CuF$_2$(pyz)(H$_2$O)$_2$ and its isotopes (see Fig.~\ref{fig1}) is similar to the chain-like compound Cu(NO$_3$)$_2$(pyz)~\cite{jamie2}. In that material electron paramagnetic resonance (EPR) suggests that the occupied $d_{x^2-y^2}$ orbitals are arranged such that one of the lobes lies along the Cu-pyz-Cu chain direction~\cite{hammar}. This, together with a lack of a strong exchange pathway between the chains leads to Cu(NO$_3$)$_2$(pyz) exhibiting extremely one-dimensional (1D) magnetic characteristics~\cite{hammar}. The situation in CuF$_2$(pyz)(H$_2$O)$_2$ is very different; the presence of the strongly electronegative F and O ions cause the $d_{x^2-y^2}$ orbitals to lie in the O-Cu-F planes diminishing the exchange along the chains~\cite{jamie2}. In addition, strong F-H-O hydrogen bonds develop between the chains that are able to efficiently mediate superexchange. The net result is that the magnetic properties of CuF$_2$(pyz)(H$_2$O)$_2$ approximate to a quasi-two-dimensional spin-$\frac{1}{2}$ Heisenberg antiferromagnet on a square lattice~\cite{jamie2}.

In this paper we show that isotopic substitution of hydrogen for deuterium in CuF$_2$(pyz)(H$_2$O)$_2$ has a sizable effect on its magnetic properties. In particular, deuteration of the water molecules acts to suppress the in-plane exchange energy without significantly changing the inter-plane exchange energy, the crystalline-electric-field local to the magnetic ions, or the crystal structure. This implies that hydrogen bonds are more efficient superexchange pathways than deuterium bonds at low temperatures. The lower mass leads to a higher zero-point vibrational energy of the hydrogen ion in the F$\cdots$H\textemdash O bond compared to that for the deuterium in the F$\cdots$D\textemdash O bond~\cite{dunitz,scheiner}. We propose that this is responsible for the observed changes in antiferromagnetic ordering temperature and critical magnetic field.

Single crystal samples of CuF$_2$(pyz)(H$_2$O)$_2$ were prepared via the recipe described in Ref~\cite{jamie2}. Four different hydrogen/deuterium isotopes were produced: one containing only hydrogen; one in which the hydrogen ions on the pyrazine ligands are replaced with deuterium; one in which the water is replaced with D$_2$O; and another containing only deuterium. For brevity the isotopes will henceforth be referred to as $h_8$, $d_4$-pyz, $d_2o$ and $d_8$, respectively.
The substitution was achieved by using commercial pyrazine-$d_4$ (Aldrich, 98 atomic \% D) and D$_2$O ($\sim95$ atomic \% D).
DC magnetization measurements were performed using a Quantum Design SQUID magnetometer with the magnetic field $B$ applied parallel to the $a$-axis. Pulsed-field magnetization experiments were performed using a 
technique described elsewhere~\cite{goddardtmagge,ho}.
Fields were provided by the 60~T
short-pulse magnets at the Nicholas Kurti Magnetic Field Laboratory, Oxford and  
the 65~T short-pulse magnets at National High Magnetic Field Labratory, Los Alamos.
The susceptometer was placed within a $^3$He 
cryostat providing temperature $T$
down to 500~mK.

\begin{figure}[t]
\centering
\includegraphics[width=8.5cm]{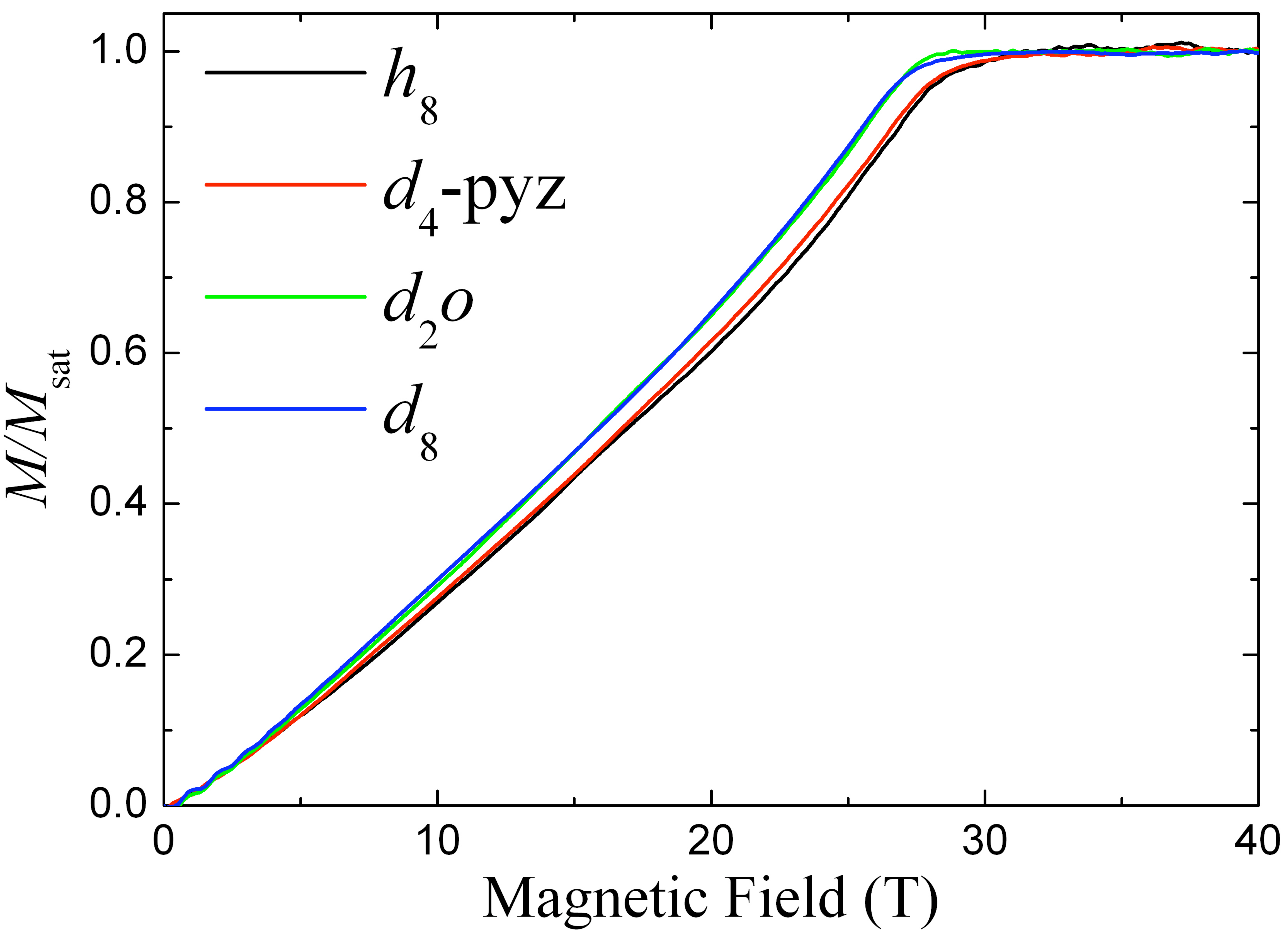}
\caption{Pulsed-field magnetization $M$ data versus magnetic field $B$ for the four different hydrogen/deuterium isotopes of CuF$_2$(pyz)(H$_2$O)$_2$. The data are taken at $T=1.5$~K with $B$ applied parallel to the $a$-axis and are normalized by the saturation magnetization $M_{\rm sat}$. $h_8$ = CuF$_2$($h_4$-pyz)(H$_2$O)$_2$, $d_4$-pyz = CuF$_2$($d_4$-pyz)(H$_2$O)$_2$, $d_2o$ = CuF$_2$($h_4$-pyz)(D$_2$O)$_2$, and $d_8$ = CuF$_2$($d_4$-pyz)(D$_2$O)$_2$.}
\label{fig2}
\end{figure}

\begin{figure}[t]
\centering
\includegraphics[width=8.0cm]{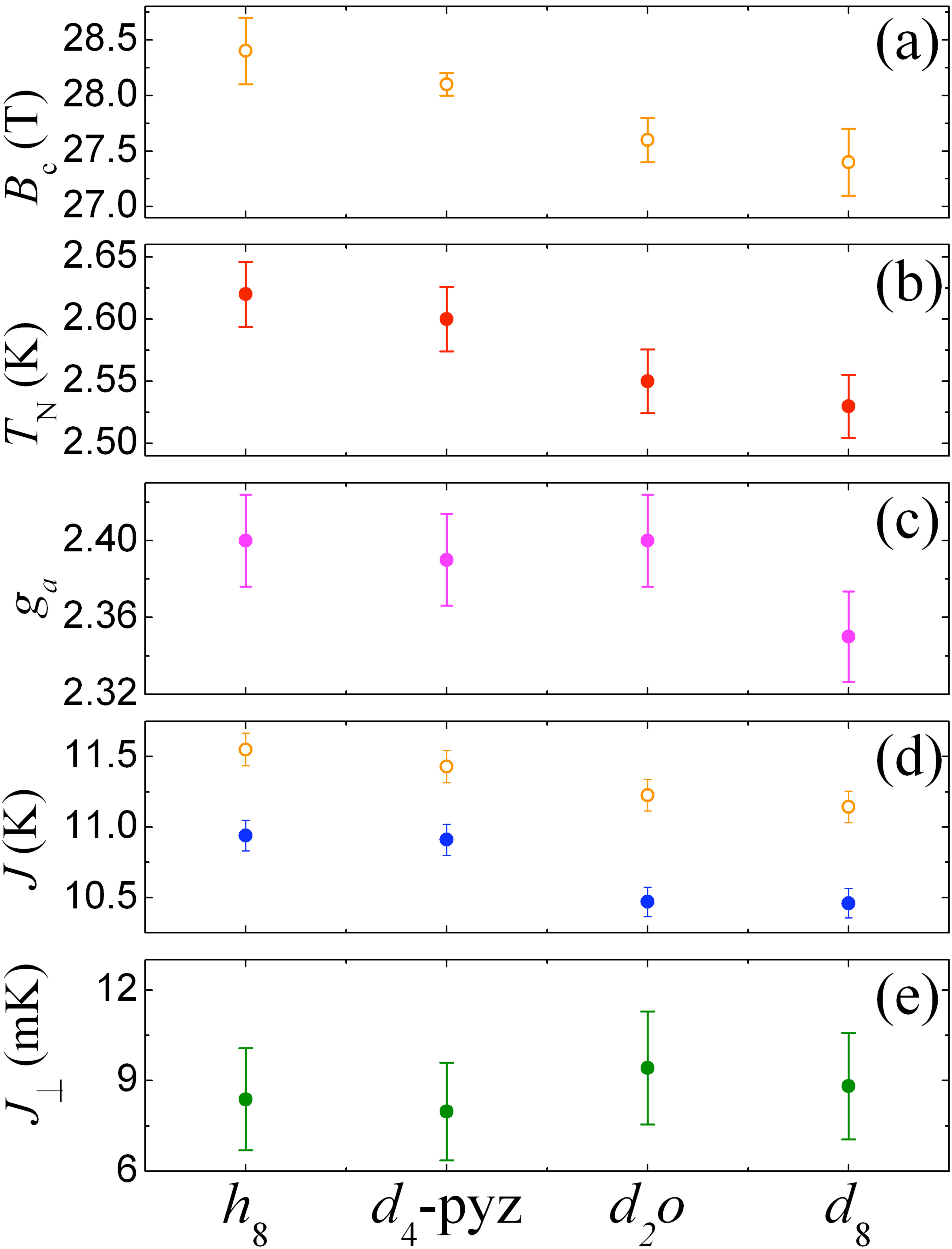}
\caption{Isotopic variation in: (a) critical magnetic field; (b) antiferromagnetic ordering temperature; (c) the $g$-factor parallel to the $a$-axis deduced from fits to $\chi(T)$ data; (d) in-plane exchange energy deduced from fits to $\chi(T)$ data (solid circles), and from critical field data (open circles); and (e) inter-plane exchange energy $J_\perp$ deduced from the expression in Ref.~\cite{yasuda}.}
\label{fig3}
\end{figure}

Typical low-temperature, pulsed-field magnetization, $M(B)$, data are shown in Fig.~\ref{fig2} for the four different isotopes with the magnetic field applied parallel to the $a$-axis. The form of the data is characteristic of quasi-2D magnets (cf. the measured and simulated magnetization curves reported in Ref.~\cite{goddardOMM}): a concave shape at low fields followed by a transition to saturation at critical field $B_{\rm c}$. The concavity of the data in Fig.~\ref{fig2} implies that these compounds have a very high degree of exchange anisotropy. Taking the critical field to be the position of the peak in the curvature of $M(B)$ we find an average $B_{\rm c}$ for each isotope using a number of datasets from several different samples at  $T\le1.5$~K. The values of $B_{\rm c}$ extracted in this way are shown in Fig.~\ref{fig3}(a). It is clear from the curves and the $B_{\rm c}$ data that deuteration of the pyrazine rings alone has very little effect on the magnetization. In contrast, deuteration of the water molecules significantly reduces the critical field. This suggests that changes in the in-plane exchange are largely responsible for the differences between the isotopes.

\begin{figure}[t]
\centering
\includegraphics[width=8.5cm]{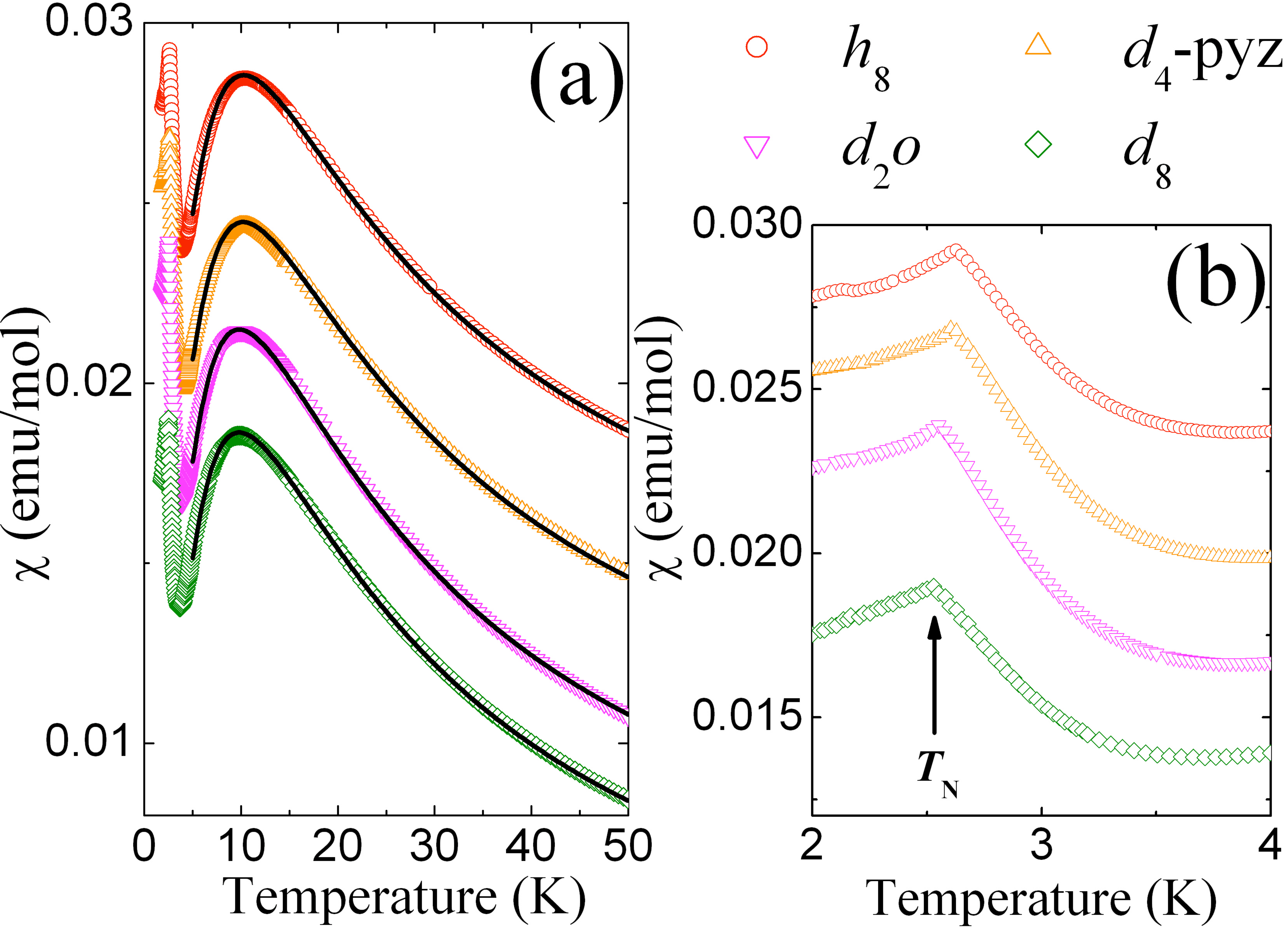}
\caption{(a) Low-field susceptibility of the four different hydrogen/deuterium isotopes of CuF$_2$(pyz)(H$_2$O)$_2$ measured with $B=0.05$~T applied parallel to the $a$-axis (points) compared to the fits to the expression from Ref~\cite{landee} (red lines). (b) Expanded view close to the ordering temperature $T_{\rm N}$. The data are offset for clarity.}
\label{fig4}
\end{figure}

A simple quasi-2D Heisenberg Hamiltonian~\cite{goddardOMM} suggests that at the critical field $g\mu_{\rm B}B_{\rm c}=4J+2J_{\perp}$~\cite{gBnote}. Because of the large exchange anisotropy (i.e. $J \gg J_\perp$) we can obtain an estimate for the in-plane exchange parameter from the $M(B)$ data. The results are shown in Fig.~\ref{fig3}(d) and make use of the $g$-factor obtained from electron paramagnetic resonance experiments on the $h_8$ isotope ($g_a=2.42\pm0.02$~\cite{goddardOMM}).
These estimates assume that the differences in $B_{\rm c}$ between the isotopes are due almost entirely to differences in the in-plane exchange parameter and that isotopic substitution of hydrogen has no significant effect on either the $g$-factor or the inter-plane exchange energy. To check the validity of these assumptions we make use of the low-field, $a$-axis susceptibility $\chi(T)$ data shown in Fig.~\ref{fig4}. 

It is seen from the figure that the susceptibilities are broadly similar for all the compounds. The antiferromagnetic ordering temperature $T_{\rm N}$ is the sharp peak observed in  $\chi(T)$ at low temperatures and its value for the four isotopes are plotted in Fig.~\ref{fig3}(b), where it is seen that they follow the same trend as the critical field data. The value of $T_{\rm N}$ agrees well with the results of other techniques~\cite{jamie2}.

The $\chi(T)$ data are fitted for $5<T<50$~K using the technique outlined in Ref.~\cite{landee}, which uses the results of theoretical predictions of the paramagnetic susceptibility of a quasi-2D spin-$\frac{1}{2}$ Heisenberg antiferromagnet with a single-exchange Hamiltonian. The fits yield values of $J$ and $g_a$.
Analysis of quantum Monte-Carlo calculations predicts that the inter-plane exchange energy $J_\perp$ for quasi-2D Heisenberg magnets is related to $J$ and $T_{\rm N}$ via the expression $|J_{\perp}| = |J|\exp\left(2.43-2.30\times|J|/T_{\rm N} \right)$~\cite{yasuda}. Thus, using the parameters obtained from the $\chi(T)$ data, the magnitudes of $J_\perp$ can be deduced for all the isotopes. The values of $g_a$, $J$ and $J_\perp$ are plotted in Fig.~\ref{fig3}(c), (d) and (e). 

As expected from the form of the $M(B)$ curves the values of $J_\perp$ are indeed small compared to $J$, leading to anisotropies of the order of $1\times 10^{-3}$. The $g$-factor for the $h8$ compound ($g_a=2.40\pm0.02$)
is in excellent agreement with that obtained from EPR experiments ($g_{\rm EPR}=2.42\pm0.02$)~\cite{jamie2}. Although the fitted values of $g_a$ vary from isotope to isotope this variation is small ($\Delta g_a/g_a \approx 2 \%$ on substitution of all the hydrogens for deuterium) and on a par with the experimental error. In addition, the small values of $J_\perp$ mean that the errors on $T_{\rm N}$, $J$ and $g_a$, once propagated through the above expression, encompass its variation across the isotopic family.  The results suggest that substitution of hydrogen for deuterium in CuF$_2$(pyz)(H$_2$O)$_2$ has very little effect on the $g$-factors or the inter-plane exchange energy. Therefore, the observed shifts in the N\'{e}el temperature and the critical magnetic field reflect the changes in the in-plane exchange energy. The variation in $J$ between isotopes with hydrogenated and deuterated water molecules is significant ($\Delta J/J \approx 4 \%$), whereas little or no difference is observed on deuteration of the pyrazine molecules alone, confirming the picture of the dominant superexchange being mediated by the hydrogen/deuterium bonds between the carbon and fluorine ions.

A shift in critical temperature is also observed on deuteration of certain layered organic molecular superconductors. Here, substitution of hydrogens on the ethylene end-groups of the bis(ethylenedithio)tetrathiafulvalene, or ET, molecule leads to an {\it increase} in the superconducting critical temperatures~\cite{kini, etisotope}. This is caused by a shortening of the C-D bond compared to the C-H bond of the hydrogenated sample, so that the spatial extent of the conducting layers is decreased and the interlayer transfer integral is slightly suppressed~\cite{etisotope,amro}. Exactly how this promotes the onset of superconductivity is not yet entirely clear, but it is suspected that reducing the dimensionality of the Fermi surface in this way enhances its susceptibility to the antiferromagnetic fluctuations believed to play a vital role in the formation of the superconducting condensate~\cite{amro}. 

CuF$_2$(pyz)(H$_2$O)$_2$ is non-metallic and so Fermi surface effects have no part in the observed isotope effect. Our preliminary neutron scattering data show that there is very little difference between the hydrogenated and deuterated structures of this material. In particular, the length of the F$\cdots$H\textemdash O hydrogen bonds and the F$\cdots$D\textemdash O deuterium bonds are identical ($d_{\rm F\cdots O} = 2.61\pm0.01$~\AA~at room temperature).  However, it is known that at low temperatures the effective ``size''  of deuterium in a chemical bond is smaller than that of hydrogen: the enhanced mass of the deuterium ion  leads to a smaller zero-point vibrational amplitude compared to the equivalent hydrogenated bond~\cite{dunitz,scheiner}. This localization leads to a reduction in the overlap of the electronic wavefunctions, a suppression of the transfer (or hopping) integral through the deuterium bond, and hence a decrease in the efficiency of the superexchange along this pathway. We suggest that this accounts for the observed shifts in $T_{\rm N}$ and $B_{\rm c}$ in the isotopes of CuF$_2$(pyz)(H$_2$O)$_2$.

This mechanism is similar to that proposed to occur in the hydrogen/deuterium isotopes of the quasi-1D Haldane-gapped antiferromagnet Ni(C$_5$H$_{14}$N$_2$)$_2$N$_3$(PF$_6$)~\cite{tsujii}. In the latter case, deuteration is believed to slightly diminish the efficiency of the secondary exchange path leading to changes in the magnetic field/temperature phase diagram. Furthermore, a reduction in the N\'{e}el temperature ($\Delta T_{\rm N}/T_{\rm N}\approx 0.6\%$) observed in La$_2$CuO$_4$ caused by substitution of O$^{18}$ for O$^{16}$ is attributed to suppression of the zero-point motion of the apical oxygen ions~\cite{zhao}. However, due to the magnitude of $J$, it is not possible to measure $B_{\rm c}$ in that compound. In contrast to those two materials, the isotope effect we have measured in CuF$_2$(pyz)(H$_2$O)$_2$ is large and seen in both $T_{\rm N}$ and $B_{\rm c}$ ($\Delta T_{\rm N}/T_{\rm N}\approx 4\%$; $\Delta B_{\rm c}/B_{\rm c}\approx 4\%$ for $h8$ and $d8$). Nevertheless, the similarity between the proposed mechanism suggests it may be a common effect in low-dimensional Heisenberg antiferromagnets. Exploitation of the effect could allow materials engineers to fine-tune exchange interactions in solid-state magnetic devices.

Finally, we note that a recent study has described how the effective sizes of bonded hydrogen and deuterium increase with temperature, and at different rates~\cite{dunitz}. The implication is that for some materials, whose magnetic properties depend crucially on superexchange through hydrogen bonds, it may be possible to observe an exchange energy that is enhanced at higher temperatures or an isotope effect whose sign changes as a function of temperature.

In summary, we have measured significant shifts in both the ordering temperature and critical magnetic field in a quasi-two-dimensional Heisenberg antiferromagnet due to isotopic substitution. We associate these shifts with differences in superexchange efficiency between hydrogen and deuterium bonds. The results highlight the fact that relatively small changes in composition can give rise to large perturbations in the properties of highly anisotropic magnetic materials; an important consideration when attempting to understand the anomalous isotope effect in the cuprate superconductors.

Work at NHMFL occurs under
the auspices
of the National Science Foundation, the State of Florida and the US Department of Energy (DoE) program ``Science at 100~T''. 
Work at ANL is
supported by the Office of Basic Energy Sciences,
DoE (contract DE-AC02-06CH11357).
Work at EWU was supported by an award from  
the Research Corporation. TL and
PAG acknowledge support from
the Royal Commission for the Exhibition of
1851 and the Glasstone Foundation respectively.
JS thanks Oxford University for the provision of
a Visiting Professorship.


\begin{thebibliography}{00}
\bibitem{gBnote}
Previous studies have found that the angle-dependence of $B_{\rm c}$ is entirely due to the anisotropy in $g$, and that the relevant energy scale for determining the magnetic properties in applied field is the product of $g$ and $B$ measured along the same direction~\cite{goddardOMM,jamie2}.
\end{thebibliography}
\end{document}